\documentclass[final]{aipproc}
\layoutstyle{6x9}

\outer\def\gtae {$\buildrel {\lower3pt\hbox{$>$}} \over 
{\lower2pt\hbox{$\sim$}} $}
\outer\def\ltae {$\buildrel {\lower3pt\hbox{$<$}} \over 
{\lower2pt\hbox{$\sim$}} $}

\begin{document}

\title{Multi-band Astronomy with {\sl LISA}}



\author{G. Branduardi-Raymont$^{*}$, G. Ramsay}{
  address={MSSL, University College London, Holmbury St Mary, Dorking, 
  Surrey, RH5 6NT, UK}
}

\author{C. Wang}{address={University of Aberdeen, Kings College, Aberdeen AB24 3UE, UK}}

\author{R. Bingham}{address={CCLRC Rutherford Appleton Laboratory, Chilton, Didcot, Oxfordshire OX11 0QX, UK}}

\author{J. T. Mendon\c{c}a}{address={Instituto Superior Tecnico, Av. Rovisco Pais, 1049-001 Lisbon, Portugal}}

\begin{abstract}
We discuss astrophysical scenarios relevant to the generation
of gravitational waves (GW) and effects expected to arise from the 
interaction of GW and electromagnetic (EM) radiation. A strong programme of 
coordinated GW and EM astrophysical studies must be established in order to 
ensure the exploitation of the full scientific potential of the LISA mission. 
We describe on-going astrophysical work, and suggest alternative approaches 
to current studies, which are relevant to these considerations.
\end{abstract}

\maketitle

\section{Introduction}

{\sl LISA} will return unprecedented data on GW sources; however, its full 
scientific potential will be realised only by matching the sources with 
astrophysical counterparts and correlating their properties over the EM 
spectrum. Different types of sources require different approaches: direct 
identification with known EM sources in some cases (like for ultra-compact 
binaries), and statistical estimation from the systems EM characteristics in 
others (such as for rates of Super Massive Black Hole, or SMBH, mergers, and 
Extreme Mass Ratio Inspirals, or EMRIs). In turn, knowledge of the EM 
properties (e.g. for the ultra-compact binaries) will be crucial in 
constructing accurate waveforms to aid {\sl LISA's} signal processing. The 
consequences of the interaction between GW and EM radiation may also have 
important implications, for instance on events such as gamma-ray bursts. It 
is clear that a large degree of synergy is needed between the GW and EM 
astrophysical communities, in order to build a strong programme of coordinated
studies targeted to the needs of {\sl LISA}. Below we discuss on-going 
astrophysical work relevant to such ideas.

\section{Ultra-compact binaries}

Ultra-compact binaries with white dwarf secondaries are predicted to be both 
numerous and strong GW sources. Indeed, these binaries will be the so-called 
{\sl Verification Binaries} for {\sl LISA}. The effects of gravitational 
radiation can be observed directly in the two candidate ultra-compact 
binaries RX J0806+15 (binary orbit 321
sec) and RX J1914+24 (569 sec). If these periods can be verified as being 
signatures of their binary orbital period, then observational work has shown 
that both systems are spinning up, i.e. their orbit is shrinking. Recent work 
indicates that RX J0806+15 should be detectable using {\sl LISA} in less 
than one 
week [1]. However, the exact mechanism which powers the EM emission in 
these candidate systems has not been settled. The proposed models fall into 
two general categories - accretion-powered and non-accretion-powered. The 
non-accretion model is that of the so-called Unipolar Inductor (UI; see 
Fig. 1, left) [2]. Although still a matter of some debate, this model perhaps 
comes closest to predicting the observational properties of RX J0806+15 and 
RX J1914+24 (e.g. [3]). The EM properties of the other known ultra-compact 
systems (orbital periods between $\sim$10 and 70 min) imply that they are 
powered by accretion.

\begin{figure}
\setlength{\unitlength}{1cm}
\begin{picture}(8,5)
\put(-6,-5.5){\includegraphics{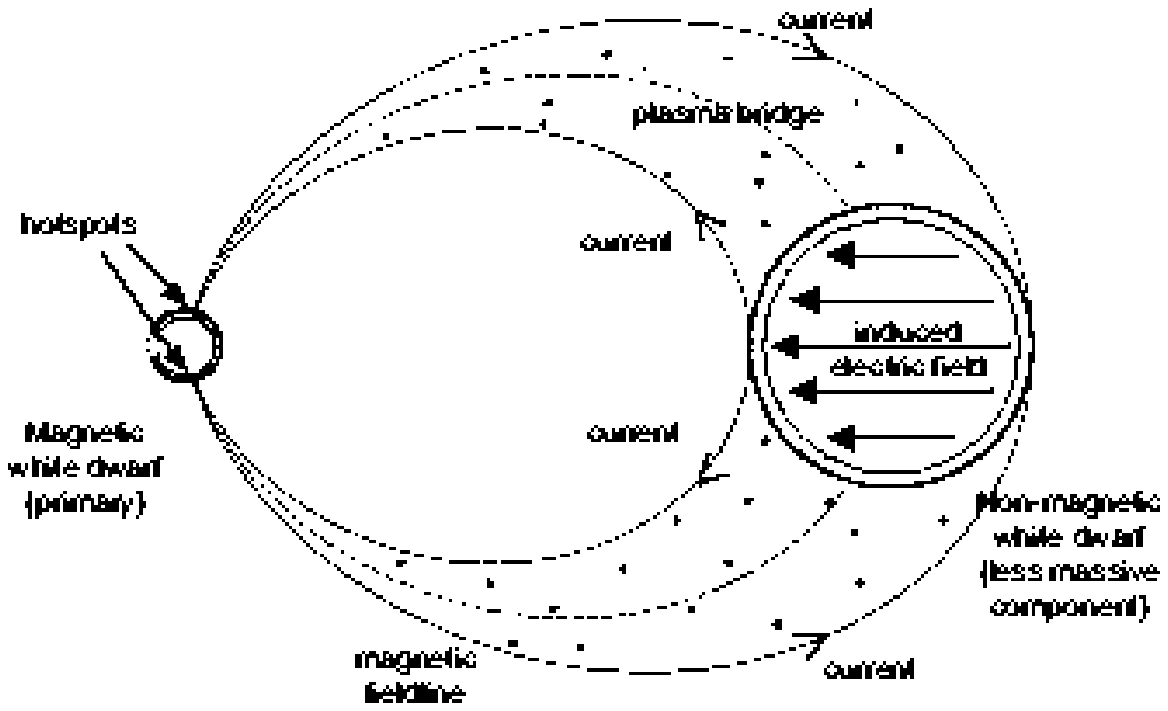}}
\put(3.5,-0.7){\includegraphics{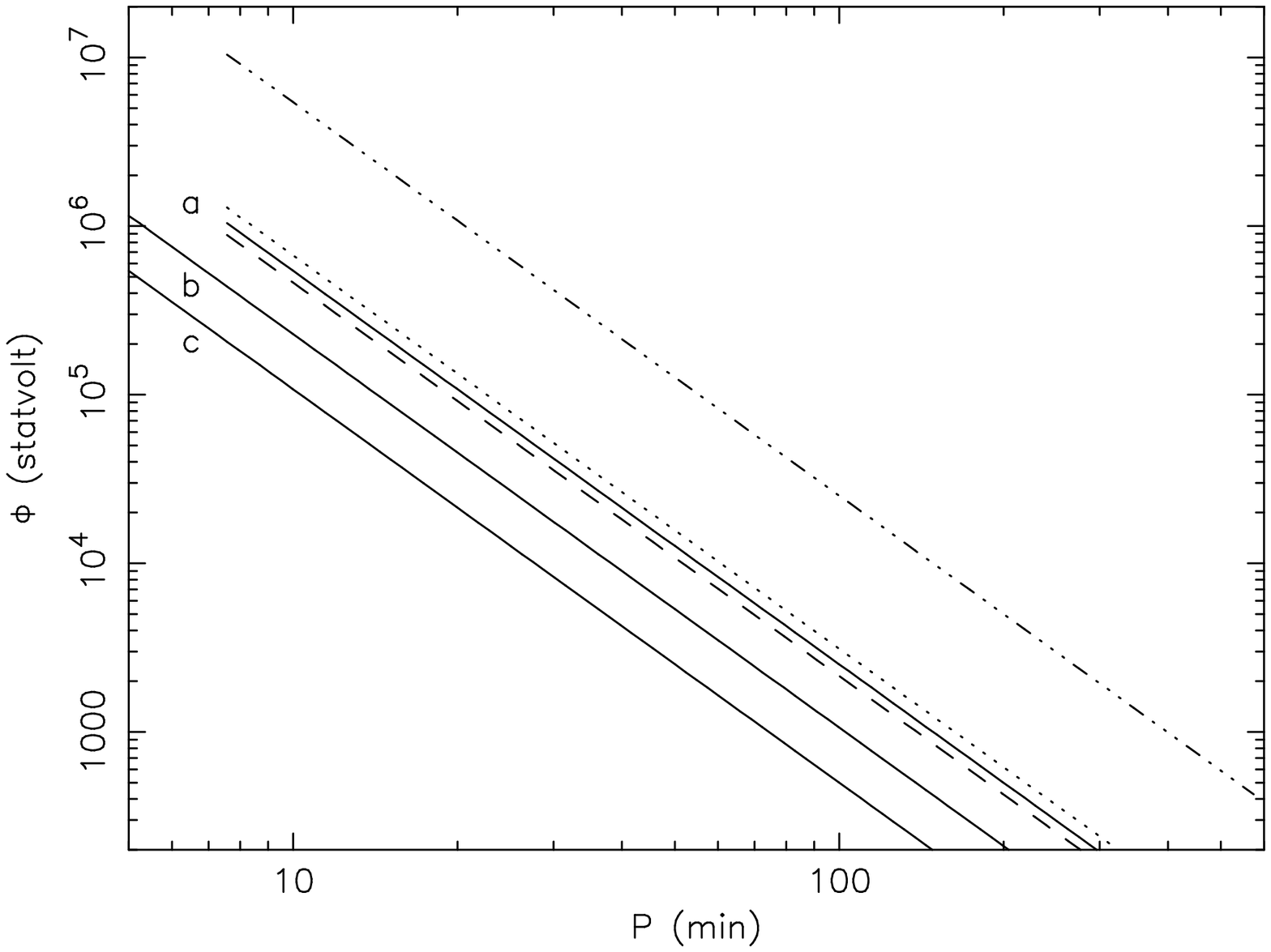}}
\end{picture}
\caption{Left: In the UI model (sketch from [2]), large electrical currents 
are driven as a conducting body (the secondary white dwarf) orbits a magnetic 
body (the primary white dwarf): these currents are dissipated at foot-points 
on the primary. The currents are so powerful that emission occurs in the 
X-ray band - the X-rays irradiate the secondary white dwarf and give rise to 
the anti-phase between X-ray and optical lightcurves [4]. Right: Power of 
electrical dissipation (solid lines) and GW (dotted) from UI systems with 9.5 
min orbital period (cf. RXJ1914+24). Curves a, b, c and d correspond to 
systems with a 0.5, 0.7, 1.0 and 1.3 solar mass primary magnetic white dwarf 
(figure from [2]).}
\label{fig1}
\end{figure}

Determining the mechanism which powers the EM emission is crucial in 
interpreting the observed spin-up rates seen in both RXJ0806+15 and 
RX J1914+24 [5, 3]. If UI is powering the EM emission, then their orbital 
evolution is determined jointly by gravitational radiation losses and EM 
interactions (Fig. 1, right) [6]. Only by knowing the relative proportion 
that each mechanism contributes to the spin-up can we correctly predict their 
GW signal.

There are reasons to expect that UI could operate in other binary systems, 
including a white dwarf orbiting a rotating black hole. This would affect 
the system's orbital evolution and hence the expected gravitational signal. 
The question is: How many systems are there? White dwarf - white dwarf 
binaries are expected to make a significant contribution to the background 
gravitational signal in the {\sl LISA} passband. Correctly modelling this 
background signal is essential for accurately predicting the sensitivity of 
{\sl LISA} observations. Theoretical models of stellar populations and binary 
evolution suggest that interacting white dwarf - white dwarf binaries are 
common place in our Galaxy. However, less than 20 of such interacting 
binaries are known. Currently it is not clear whether this discrepancy is 
due to inadequacies in the theoretical models or that many more interacting 
binaries await discovery.

A number of projects are on-going to test this question by searching for new 
systems. The SPY project (Sn Ia Progenitor surveY), for instance, searches 
for radial velocity variations in a large sample of faint white dwarfs [8]. 
The aim is to determine the orbital period distribution of such binaries and 
also test whether these systems can give rise to type Ia supernovae. A 
different approach is provided by RATS (RApid Temporal Survey) [9]. This 
study searches for stellar sources whose intensity varies on short periods, 
from a few minutes to periods longer than several hours. A pilot field 
included the binary RX J0806+15 (321 sec orbital period): such kind of system 
can easily be detected (Fig. 2). RATS is on-going but initial results suggest 
that theoretical models significantly overestimate the space density of 
interacting white dwarf - white dwarf binaries.

\begin{figure}
\setlength{\unitlength}{1cm}
\begin{picture}(8,5)
\put(-1.5,-6.7){\includegraphics{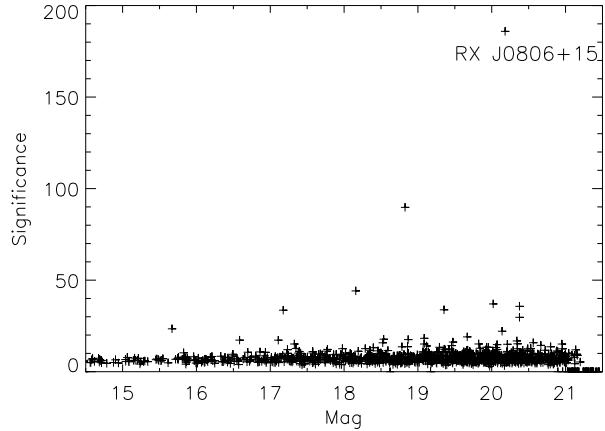}}
\end{picture}
\caption{Results from RATS (figure from [9]) where objects are selected on 
the basis of their variability characteristics: higher significance values 
imply strong variability. RX J0806+15 is clearly identified as being highly 
variable.}
\label{fig2}
\end{figure}

\section{Super Massive Black Holes (SMBHs) as GW sources}

Predictions of event rates of both, SMBH binary mergers and EMRIs are heavily 
dependent on the universal mass density distribution of SMBHs. A strong 
correlation has been established between SMBH mass and galactic velocity 
dispersion (and a weaker one with luminosity of the host galaxy's stellar 
bulge). The correlation has been used to derive the local SMBH mass density. 
The AGN luminosity function as a function of redshift, i.e. the 
representation of their evolution with cosmic time, traces the accretion 
history of the BH and gives a measure of the accreted mass density, and 
ultimately the mass distribution of SMBHs [10]. Such calculations have been 
based so far on the luminosity function of optically bright QSOs.

However, a discrepancy exists between the strong optical evolution of AGN at 
z\ltae 2, and the X-ray luminosity function which peaks at z$\sim$1. The 
existence of a significant number of absorbed AGN making up a large fraction 
of the entire population may explain this inconsistency. Deep pencil-beam 
X-ray surveys of AGN appear to indicate that the amount of obscuration is 
strongly luminosity-dependent, with the fraction ($>$ 75\%) of obscured AGN 
being larger at low luminosities.

On the other hand, XMM-Newton large area surveys [11] show that the pattern 
of absorption in AGN is independent of both redshift and luminosity, with 
obscured AGN being $\sim$3 times more populous than un-obscured ones at all 
redshifts and luminosities. This is illustrated by the presence of absorbed 
objects (filled rectangles) in a $L_{X} - z$ plot (Fig. 3) [12].

\begin{figure}
\setlength{\unitlength}{1cm}
\begin{picture}(8,10)
\put(-0.5,0.0){\includegraphics{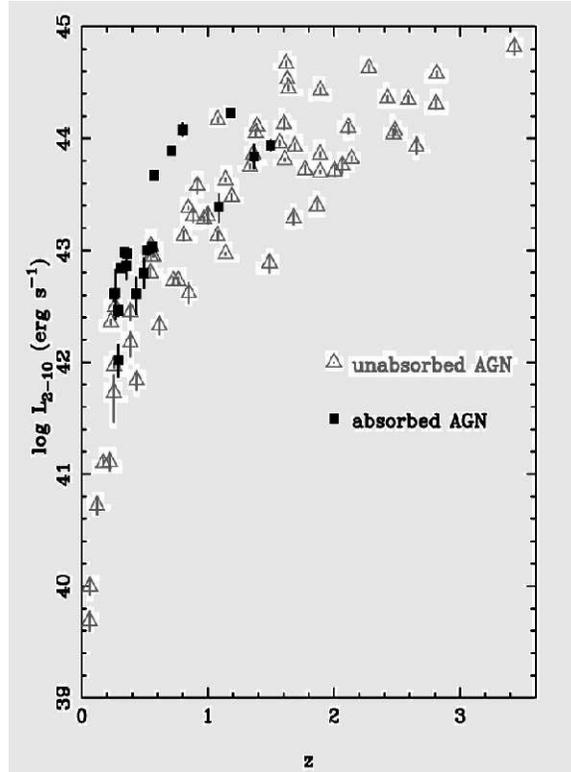}}
\end{picture}
\caption{2--10 keV intrinsic rest-frame luminosities as a function of 
redshift for AGN detected with {\sl XMM-Newton} in the 13H deep field. Sources 
which show absorption in their X-ray spectra are indicated by filled 
rectangles, while those without significant X-ray absorption are shown by 
triangles.}
\label{fig3}
\end{figure}

Moreover, the space density of X-ray selected AGN has been found to be  up to 
a factor of 40 larger than that of bright optically selected QSOs [13]. These 
results require that estimates of the SMBH mass distribution be reconsidered. 

Our knowledge of the demographics and evolution of AGN, and thus the accuracy 
of the SMBH mass distribution, is bound to improve further as we plan to 
combine the wide angle, deep {\sl XMM-Newton} surveys with mid-infrared 
(Spitzer) 
imaging surveys, which have the potential of revealing the most distant and 
heavily absorbed AGN.

\section{Gamma-ray bursts (GRB's)}

The detection by {\sl Swift} of more than 120 GRBs so far (mid 2006) has 
already revolutionarised our view of these most energetic phenomena, which 
are thought to be associated with the coalescence of neutron stars and 
black holes, and thus GW production. As statistics improve with more bursts 
being detected, the characterisation of larger samples of short (< 2 sec) 
and long bursts will provide an estimate of the relative frequency of the 
different types of mergers, be coalescent neutron stars, black hole mergers 
or hypernova events.

\section{Photon energy up-shift by plasma waves induced by GW from a compact
source}

The highly non-linear nature of GW at source results in the coupling to other 
wave modes such as plasma waves. The generation of these wave modes causes an 
attenuation of the GW. For strong GW burst models the Bondi-Sachs metric has 
been used to evaluate the non-linear modification of the effective refractive 
index. These models show that photons and high-energy particles can 
experience significant energy shifts by 'surfing' on the plasma waves [14]. 
This effect may have important implications on gamma-ray events such as GRBs 
and causal GW and EM observations.

\section{Acknowledgements}

The MSSL authors thank M. J. Page for producing Fig. 3 and for useful 
discussions. Figures 1 and 2 are re-printed with permission 
from Mon. Not. R. Astron. Soc., Blackwell Publishing.

\section{References}

1. A. Stroeer and A. Vecchio, in Classical and Quantum Gravity, Proceedings 
of the 10th Gravitational Wave Data Analysis Workshop, submitted 
(2006).\newline
2. K. Wu et al., Mon. Not. R. Astron. Soc. 331, 221-227 (2002).\newline
3. G. Ramsay, M. Cropper and P. Hakala, Mon. Not. R. Astron. Soc. 367, 
L62-L65 (2006).\newline
4. G. Ramsay et al. Mon. Not. R. Astron. Soc. 311, 75-84 (2000).\newline
5. P. Hakala, G. Ramsay and K. Byckling, Mon. Not. R. Astron. Soc. 353, 
453-456 (2004).\newline
6. S. Dall'Osso, G. L. Israel and L. Stella, A\&A 447, 785-796 (2006).\newline
8. R. Napiwotzki et al., Astron. Nachr. 322, 411-418 (2001).\newline
9. G. Ramsay and P. Hakala, Mon. Not. R. Astron. Soc. 360, 314-321 
(2005).\newline
10. Q. Yu and S. Tremaine, Mon. Not. R. Astron. Soc. 335, 965-976 
(2002).\newline
11. T. Dwelly and M. Page, Mon. Not. R. Astron. Soc., in press 
(astro-ph/0608479) (2006).\newline
12. M. J. Page et al., Mon. Not. R. Astron. Soc. 369, 156-170 (2006).\newline
13. G. Hasinger, T. Miyaji and M. Schmidt, A\&A 441, 417-434 (2005).\newline
14. C. Wang et al., in preparation (2006).\newline

\end{document}